**Tunable electron scattering mechanism in plasmonic SrMoO₃ thin films**

*Rahma Dhani Prasetiyawati[1], Seung Gyo Jeong[1], Chan-Koo Park[1,2], Sehwan Song[3], Sungkyun Park[3], Tuson Park[1,2], and Woo Seok Choi[1]\**

R. D. Prasetiyawati, S. G. Jeong, and W. S. Choi
[1]Department of Physics, Sungkyunkwan University, Suwon, 16419, Korea
E-mail: choiws@skku.edu

C. K. Park and T. Park
[2]Center for Quantum Materials and Superconductivity (CQMS), Sungkyunkwan University, Suwon 16419, Korea

S. Song and S. Park
[3]Department of Physics, Pusan National University, Busan 46241, Korea

\*E-mail: choiws@skku.edu

4$d$ transition metal perovskite oxides serve as suitable testbeds for the study of strongly correlated metallic properties. Among these, SrMoO₃ (SMO) exhibits remarkable electrical conductivity at room temperature. The temperature-dependent resistivity ($\rho(T)$) exhibits a Fermi-liquid behavior below the transition temperature $T^*$, reflecting the dominant electron-electron interaction. Above $T^*$, electron-phonon interaction becomes more appreciable. In this study, we employed the power-law scaling of $\rho(T)$ to rigorously determine the $T^*$. We further demonstrate that the $T^*$ can be modified substantially by ~40 K in epitaxial thin films. It turns out that the structural quality determines $T^*$. Whereas the plasma frequency could be tuned by the change in the electron-electron interaction via the effective mass enhancement, we show that the plasmonic properties are more directly governed by the electron-impurity scattering. The facile control of the electron scattering mechanism through structural quality modulation can be useful for plasmonic sensing applications in the visible region.

## 1. Introduction

In the correlated electron systems, several scattering mechanisms govern their transport properties. As summarized in Fig. 1(a), the electron-impurity scattering influences the electronic properties independent of the temperature ($T$) reflected in the residual resistivity $\rho_0$. Meanwhile, most of the correlated metallic states at low $T$ have been successfully described by the Fermi-liquid (FL) theory [1-3]. The electron-electron scattering in FL metals at a low $T$ determines the electronic transport behavior, reflected by the $T^2$-dependence of the resistivity ($\rho$), (Fig. 1(a)) [4]. When $T$ increases above the transition $T$ ($T^*$), $\rho(T)$ departs from the $T^2$ behavior as the electron-phonon coupling becomes the primary scattering mechanism. In general, $\rho(T)$ can be defined by the following equation,

$$\rho = \rho_0 + AT^\alpha, \tag{1}$$

where $A$ is the $T$-dependent coefficient and the $T$-exponent $\alpha$ characterizes the type of scattering contributing to the transport properties of the system. $\alpha = 2$ indicates the FL behavior with a dominant contribution from electron-electron scattering.

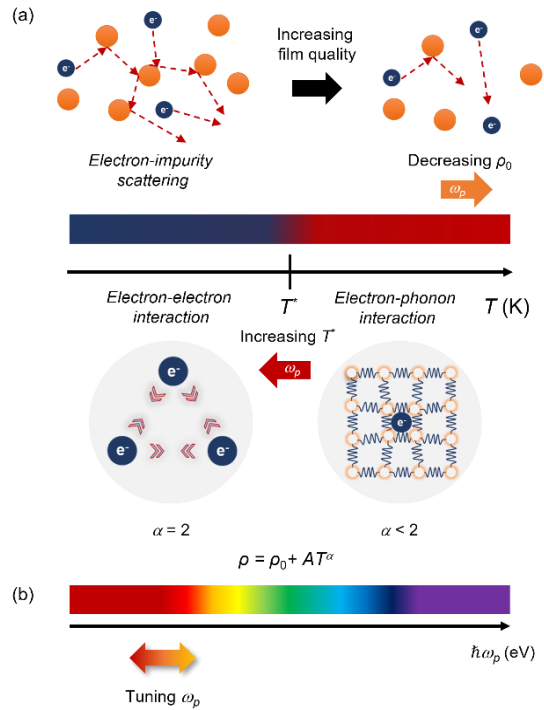

**Figure 1.** Schematic diagram of (a) the *T*-independent electron-impurity scattering represented by the residual resistivity ($\rho_0$) and *T*-dependent evolution of the scattering mechanisms in Fermi-liquid metals that influences the transport properties of SrMoO$_3$. While decreasing $\rho_0$ is expected to shift plasma frequency ($\omega_p$) to higher energy, increasing $T^*$ will oppositely lower $\omega_p$. (b) $\omega_p$ can be tuned through the modulation of the degree of electron-electron correlation and electron-impurity scattering.

4*d* transition metal perovskite oxides are suitable model systems for realizing and investigating strongly correlated metallic behaviors of FL phases. The complex interactions between various degrees of freedom in the 4*d* transition metal oxides play an important role in the revelation of emerging correlated electronic phenomena such as superconductivity [5], non-FL behavior [6-8], metal-insulator transitions [9-11], spin-orbit interaction-driven phases [12, 13], and the anomalous Hall effect [14, 15]. The relevant electronic correlations of 4*d* ions include the on-site Coulomb interaction, bandwidth, crystal field, Hund's coupling, and spin-orbit interaction, widening the strongly correlated regime beyond that of the conventional 3*d* transition metal oxides [16-19]. The more extended nature of the 4*d* orbitals results in a large bandwidth and small screened Coulomb interaction compared with the 3*d* orbitals [19, 20]. 4*d* perovskite oxides composed of Nb, Mo, Ru, and Rh are mostly metallic because their $t_{2g}$ shells are not half-filled. This is favorable for the application of metallic electrodes in oxide electronics, including potential plasmonic device applications [21, 22].

Among conducting 4*d* perovskite oxides, SrMoO$_3$ (SMO) is largely underexplored in terms of its electronic transport mechanism. This is notwithstanding its exceptionally small room-*T* $\rho$ and its potential application for transparent conducting electrodes [4, 23, 24]. The small room-*T* $\rho$ results from the high carrier concentration (*n*), in the order of ~$10^{22}$ cm$^{-3}$. The metallic behavior of SMO originates from the strong hybridization between the O 2$p$ and Mo 4$d$ $t_{2g}$ states near the Fermi level. The interband transition between these two orbitals occurs at a relatively high photon energy of ~4 eV, ideally giving rise to transparency in the visible to the ultraviolet region [25, 26]. In conjunction with thermal stability up to 450 °C in air and up to 700 °C in a vacuum, this property renders SMO suitable for plasmonic applications [27, 28]. Unlike its ruthenate counterpart (SrRuO$_3$, which is ferromagnetic), SMO exhibits Pauli paramagnetic behavior with a $T$-independent magnetic susceptibility [29, 30]. This may eliminate complications involving magnetism and corroborates the influence of electron-electron interactions on the optoelectronic properties of SMO. In addition, the renormalization of the quasiparticle peak in the photoemission spectrum of SMO is ascribed to the effect of Hund's coupling on the partially filled $t_{2g}$ bands [18]. On the other hand, the Kadowaki–Woods ratio (A/$\gamma^2$ = 1.1 × 10$^{-6}$ $\mu\Omega$·cm/(mJ/K mol)$^2$) of the SMO single crystal from the specific heat measurement corresponds to that of heavy-fermion compounds indicative of the FL behavior [4].

The FL behavior of SMO was observed in its $\rho(T)$ below the $T^*$. The two reported $T^*$ values of SMO in the literature, 140 and 85 K for bulk single-crystal and thin-film SMO on SrTiO$_3$ (STO) substrate, respectively, exhibit a substantial inconsistency [4, 23]. This is partly because different methods were used to characterize $T^*$. For example, the $T^*$ values have been obtained by plotting $\rho$ as a function of $T^2$ and then selecting the point at which $\rho$ deviates from a linear fit. In contrast, the exponent $\alpha$ can be obtained more quantitatively by employing the power-law scaling of $\rho(T)$, thereby rigorously determining the $T^*$ value for the FL behavior. Furthermore, the non-FL behavior, e.g., $\rho$ with $T^{1.5}$-dependence, above the $T^*$ can be identified using the power-law analysis. Because the FL behavior fundamentally determines the transport properties of SMO, it is important to accurately identify and effectively tune $T^*$, for the facile adaptation of the functionality.

In this study, we demonstrated that $T^*$ can be largely tuned by ~40 K within SMO epitaxial thin films, indicating the feasibility of modifying the electron scattering mechanism through structural quality control. More specifically, the $T^*$ value was modified by employing different laser fluences during the pulsed laser epitaxy (PLE) growth of thin films. Despite the overall similarity of the perovskite lattice, chemical, and general transport behaviors across all the films, the selection of the laser fluence effectively modified the kinetic energy of the ablated species, governing the microscopic structural quality [31, 32]. The degree of electron correlation reflected in $T^*$ was demonstrated to be closely linked to the film quality represented by $\rho_0$. Based on these tunable electron interactions, effective control of plasma frequency ($\omega_p$) in the visible region was demonstrated by modulating $n$ and/or effective mass ($m^*$), as shown in Fig. 1(b), according to the equation,

$$\omega_p = \frac{e}{2\pi}\sqrt{\frac{n}{\varepsilon_0 \varepsilon_r m^*}}, \qquad (2)$$

where $e$ is the elementary charge, $\varepsilon_0$ is the vacuum permittivity, and $\varepsilon_r$ is the relative permittivity of the material. $\omega_p$ governs the plasmonic behavior of the material in conjunction with the epsilon-near-zero (ENZ) energy. The ENZ energy is the crossover energy where the real part of the dielectric function ($\varepsilon_1(\omega)$) changes sign as a function of the photon energy, indicating the change of behavior from plasmonic to dielectric. In Fig. 1(a), an increase in the electron-electron interaction is expected to decrease $\omega_p$ through the enhancement of $m^*$. Meanwhile, with decreasing $\rho_0$, $\omega_p$ is expected to shift to the higher energy with increase in $\tau$. We show that the electronic and plasmonic behavior of SMO thin films is primarily governed by the $T$-independent electron-impurity scattering overcoming the contribution from the electron-electron interaction below $T^*$ as shown in Fig. 1(b). The controllable plasmonic behavior of SMO thin films achieved by collectively tailoring the electron scattering via the film quality modulation can potentially be applied to plasmonic sensing, adding alternative material systems in addition to noble metals such as silver and gold [33-35].

## 2. Experimental Methods

High-quality SMO thin films were deposited on (110) GdScO$_3$ (GSO) substrates using PLE. Before the deposition process, the surfaces of the GSO substrates were treated using buffered HF and then annealed at 1100 °C for 6 h. A ceramic target of SrMoO$_4$ was used. The deposition was performed in a high-vacuum condition ($P < 5 \times 10^{-7}$ Torr) to reduce the Mo$^{6+}$ oxidation state of the target to the Mo$^{4+}$ state of the targeted perovskite phase with a substrate temperature of 600 °C. An excimer (KrF) laser (248 nm; IPEX 868, LightMachinery) with a repetition rate of 2 Hz was used for the target ablation. The energy from the laser is used to ablate the target material and to form a plume. The initial velocities of the ablated materials depend on the laser fluence, and it will influence how the ablated species propagate and crystallize on top of the substrate. Additionally, the bond-formation energy difference between the cations might also lead to preferential deposition potentially causing the deviation in the cation stoichiometry. The compositional deviation would affect the lattice parameters and in turn, govern the microscopic structural quality of the deposited thin films. For example, in STO, more Ti species are ablated at higher laser energy creating Sr-deficient films [36]. To achieve the targeted plasmonic properties, we need to carefully search for the optimum laser fluence for the film deposition. Here, the laser fluences were varied from 0.48 to 0.86 J·cm$^{-2}$ by systematically modifying the laser energy and aperture size.

The structural quality of the SMO thin films was assessed using high-resolution XRD (X'Pert PRO, Malvern Panalytical) with Cu K$\alpha$ radiation. The thicknesses of the films were in the range of 43 to 62 nm and were determined by using X-ray reflectivity (XRR). The chemical composition of the SMO thin films and the oxidation states of all the elements were studied using XPS (AXIS Supra+, Kratos Analytical Ltd.) at room-$T$. For XPS, an Al-K$\alpha$ monochromator x-ray source ($h\nu$ = 1486.6 eV) with a step size of 1 eV at pass energy of 200.0 eV with a spot size of 400 $\mu$m was used. Prior to the XPS measurements, the films were etched using Ar$^+$ ions with an etching energy of 1 kV and an etching area of 1.5 × 1.5 mm$^2$ to measure the inner part of the film. The electrical properties measurements were performed using a Teslatron™PT (Oxford Instruments) and physical property measurement system (PPMS-Dynacool, Quantum Design). $\rho$ as a function of $T$ was collected from 300 K to 1.9 K using the van der Pauw configuration with In electrodes and Au wires on the 5 × 5 mm$^2$ film. The optical measurements were performed using spectroscopic ellipsometry (M-2000® Ellipsometer, J. A. Woollam, Co., Inc.). The photon energy used was in the range of 0.74 – 6.50 eV with incident angles of 55°, 60°, and 65°. We employed a two-layer model analysis (an SMO thin film on a GSO substrate) to obtain physically reasonable spectroscopic dielectric functions along the in-plane direction of the film using WVASE® software. Five fitting parameters were used: $\varepsilon_{1,\text{sub}}(\omega)$, $\varepsilon_{2,\text{sub}}(\omega)$, $t_{\text{film}}$, $\varepsilon_{1,\text{film}}(\omega)$, and $\varepsilon_{2,\text{film}}(\omega)$ which correspond to the real and imaginary parts of the dielectric function of the substrate, the film thickness, and the real and imaginary parts of the dielectric function of the film, respectively.

## 3. Results and Discussion

The epitaxial SMO thin films were stabilized in tetragonal crystal structures on (110) GSO substrates (Fig. S1). To achieve high-quality SMO (bulk cubic lattice parameter, $a_c$ = 3.98 Å) thin films, GSO (pseudocubic lattice parameter, $a_{pc}$ = 3.96 Å) substrates with a small lattice mismatch of −0.20% were chosen. Fig. 2 shows the lattice structures of SMO thin films grown with various laser fluences (0.48 – 0.86 J·cm$^{-2}$). The X-ray diffraction (XRD) $\theta$-$2\theta$ scans verified the growth of the high-quality SMO thin films as shown in Fig. 2(a). In particular, whereas pulsed laser deposition frequently induced the formation of SrMoO$_4$ impurities [25], our growth at a high vacuum ($< 5 \times 10^{-7}$ Torr) resulted in phase-pure SMO thin films regardless of the laser fluence used. The presence of Pendellösung fringes around the film peaks further indicated a smooth film/substrate interface and the film surface. The rocking curve scan of the (002)$_{pc}$ peaks for the GSO substrate and SMO film grown are shown in Fig. 2(b) and 2(c), respectively, for the thin film grown at a laser fluence of 0.61 J·cm$^{-2}$. The full-width-at-half-maximum (FWHM) values of the $\omega$-scan peaks were obtained in the range of 0.050° – 0.068° indicating a high crystallinity. The XRD reciprocal space maps around the (103)$_{pc}$ reflection of the GSO substrate (Fig. 2(d)) confirm that all the epitaxial SMO thin films were fully strained to the substrates. The in-plane and the out-of-plane lattice parameters were 3.97 and ~4.03 Å, respectively, confirming the tetragonal symmetry of the thin films. The XRD results indicated that the lattice structures of the thin films were consistent across the laser fluences used. In addition, the X-ray photoelectron spectroscopy (XPS) results implied that the atomic concentrations (stoichiometry) of all the films were similar within the large error bar, supporting the XRD results (Supplementary Information S1, Fig. S2, and Table S1). Nevertheless, the XPS results may suggest the presence of Sr vacancies in the SMO thin films which might act as scattering centers.

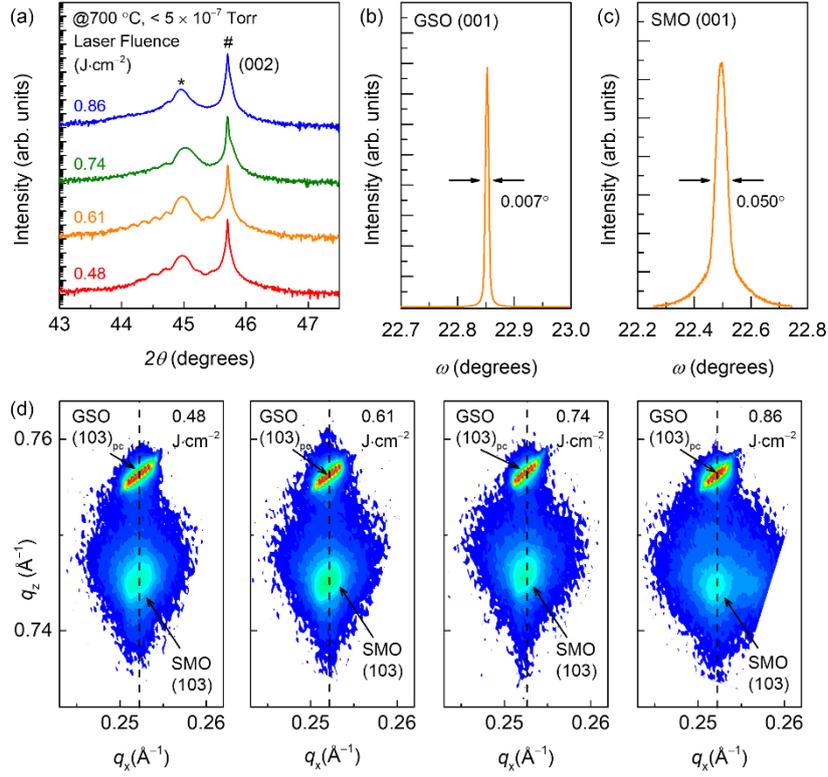

**Figure 2.** (a) XRD $\theta$-$2\theta$ scans of the SMO epitaxial thin films grown on GSO substrates. * and # indicate film and substrate peaks, respectively. XRD rocking curve scans of (b) GSO substrate and (c) SMO film grown using a laser fluence of 0.61 J·cm$^{-2}$. (d) XRD reciprocal space maps for the (103) peaks of SMO films grown using laser fluences of 0.48, 0.61, 0.74, and 0.86 J·cm$^{-2}$. The in-plane lattice constants of all the films are the same as that of the GSO substrates, as indicated by the dashed line.

The global $T$-dependent transport properties of the SMO thin films also exhibited a consistent behavior, followed by their structural properties and chemical stoichiometry. This suggests the robust growth of the thin films across a wide laser fluence window. All the SMO films exhibited metallic $\rho(T)$ behavior, as shown in Fig. 3. We also display the solid and dotted fitting lines for $T^2$- and $T^{1.5}$ dependences, respectively. These indicate good fits for the low- and high-$T$ regions, respectively (Fig. S3 and S4). However, the overlapping region for the two fits is significantly wide, hindering the accurate determination of $T^*$, as were the cases in previous studies [4, 23].

To obtain more rigorous values of $T^*$, we performed power-law exponent analysis and calculated the exponent $\alpha$ across all $T$s, i.e.,

$$\alpha(T) = \frac{d(\ln\rho - \rho_0)}{d(\ln T)}. \qquad (3)$$

The extracted $\alpha(T)$ values are shown on the right axes in Fig. 3. As anticipated, $\alpha(T)$ increases as $T$ decreases. We define $T^*$ as the $T$ when $\alpha = 2$ marked by the red horizontal dashed line, which corresponds to the FL behavior. We also use vertical dashed lines in Fig. 3 to show the position of $T^*$ indicating the change from FL to non-FL behavior with increasing $T$.

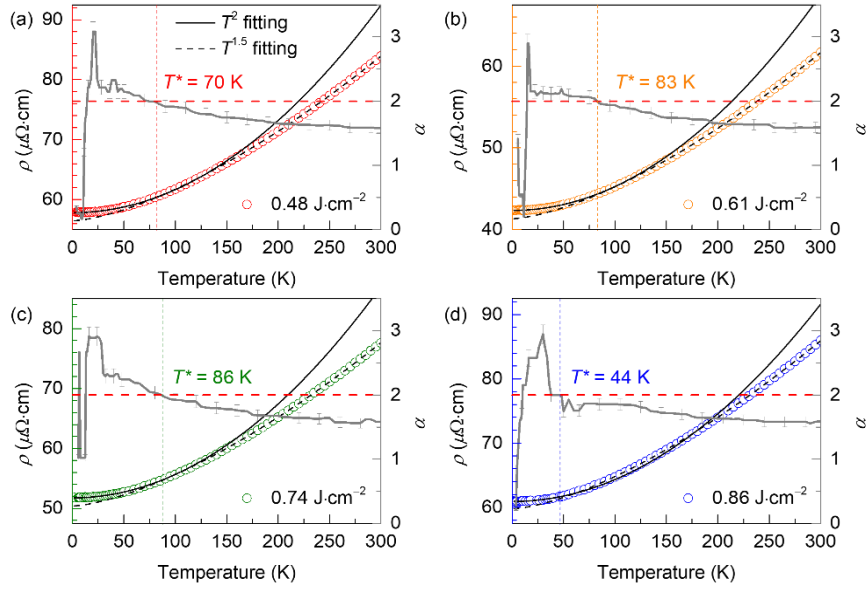

**Figure 3.** (Left) $\rho(T)$, fitting lines using $T^2$-dependence below $T^*$ (black solid lines) and $T^{1.5}$-dependence above $T^*$ (black dashed lines), and (right) the power-law exponent $\alpha$ values as a function of $T$ for the SMO epitaxial thin films grown with laser fluences of (a) 0.48, (b) 0.61, (c) 0.74, and (d) 0.86 J·cm$^{-2}$. The $T^*$ values are marked by vertical dashed lines to indicate the transition from the FL to the non-FL behavior. The red horizontal dashed line marks $\alpha = 2$ indicating the FL behavior.

The obtained $T^*$ values exhibited a large variation in the range of 44 – 86 K despite the consistent metallic behavior. This suggests that subtle changes in the growth conditions can have a significant impact on the dominant electron interaction in the FL system. Compared with the reference values of $T^*$ (note that the values may not have been accurately obtained, as discussed in the Introduction), it was higher (lower) than that of the thin film (bulk single crystal) [4, 23]. The overall reduced $T^*$ for the thin films compared with that for the bulk single crystal indicated the suppressed electron-electron interaction in the thin film due to possibly enhanced disorder.

From the power-law scaling analysis, a non-FL behavior was observed above the $T^*$ indicated by the broad downturn of $\alpha$ toward $\alpha = 1.5$, (Fig. 3(a)-(d)). Earlier, $T^{1.5}$-dependence was observed in 4$d$ perovskites such as SrRuO$_3$ (20–160 K) and CaRuO$_3$ (1.5–10 K) [37-39]. These are interpreted as fluctuations in the magnetically ordered phase. However, SMO does not exhibit long-range magnetic ordering, as evidenced by the paramagnetic behavior in Hall resistance measurements, without any anomalous Hall effect (Fig. S5(a)). In addition, the non-FL phase of the SMO thin film was significantly wide in the $T$-range, ranging from $T^*$ (44 – 86 K) to room-$T$ or higher. The absence of magnetic ordering in SMO may have resulted in a more extensive non-FL behavior. It is noteworthy that the emergence of non-FL behavior in SMO above $T^*$ can be influenced by the effect of Hund's coupling. Previously, the effect of Hund's coupling has been observed in SMO in terms of the quasiparticle peak renormalization [18]. Hund's coupling emerges in multiorbital materials with partially filled $t_{2g}$ shells, such as most correlated 4$d$ transition metal oxides. This coupling results in a strong electronic correlation competing with the Mott phase and lowers the quasiparticle coherence scale which in turn makes the metallic state to be more correlated [19, 20]. Here, $T^*$ also serves as the FL scale, indicating the coherent regime of SMO. The incoherent regime above $T^*$ is manifested in the non-$T^2$-dependent $\rho$ that is considered to be the prominent effect of electron-phonon coupling.

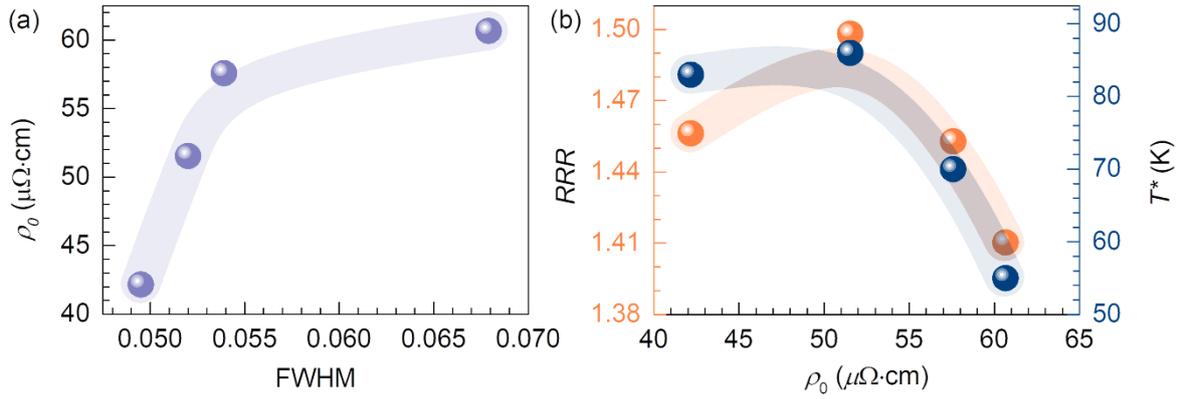

**Figure 4.** (a) $\rho_0$ of SMO thin films as a function of the FWHM value from the XRD rocking curves. (b) Residual resistivity ratio (RRR, left) and $T^*$ (right) of SMO thin films as functions of $\rho_0$.

The degree of electron-electron interaction in SMO thin films reflected by the $T^*$ values is determined sensitively by the minuscule difference in the lattice coherence. For example, Fig. 4(a) shows the residual resistivity $\rho_0$ (the $\rho$ value at 0 K) values of the SMO thin films as a function of the quantified lattice coherence, which is the FWHM value in the XRD rocking curve scans. As expected, $\rho_0$ decreases systematically by 30% as the FWHM value in the XRD rocking curve scans decreases from ~0.07 to ~0.05, by 29%. A marginal improvement in the film crystallinity can effectively alter its transport properties, particularly in the regions with good lattice coherence (the decreasing slope is steeper when the FWHM value is smaller). The overall sample quality can also be assessed from the residual resistivity ratio (RRR), i.e., $\rho_{300 K}/\rho_0$. The RRR and $T^*$ values of all the films are both inversely proportional to $\rho_0$ as summarized in Fig. 4(b). $\rho_0$ corresponds to the degree of electron-impurity scattering, which is related to the number of defects acting as the scattering centers. Because RRR is tied strongly to $T^*$, the film quality reflected in $\rho_0$ evidently affects the dominant scattering mechanism that contributes to the electrical transport properties of SMO. The obtained RRR values are comparable to that of the previous SMO thin film which is 1.5. However, these differ from the RRR value of the single crystal SMO by an order of magnitude [4, 23, 40, 41]. This is further corroborated by the higher $T^*$ value for single-crystal SMO ($T^* = \sim 120$ K) than an SMO thin film on STO substrate ($T^* = \sim 80$ K) (Supplementary Information S4 and Fig. S7).

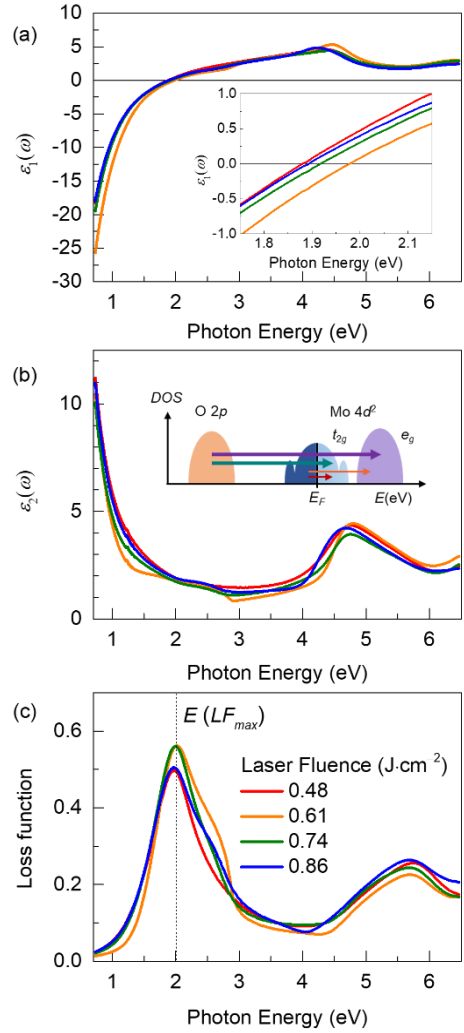

**Figure 5.** (a) Real $\varepsilon_1(\omega)$ and (b) imaginary parts $\varepsilon_2(\omega)$ of dielectric functions of SMO thin films at room-$T$. The inset of (a) shows the enlarged part near ENZ, and the inset of (b) shows the schematic band structure of the SMO. (c) The loss function of SMO thin films.

The electron scattering mechanism control also plays an important role in obtaining the desired plasmonic properties because it significantly affects the plasmon-damping process. For plasmonic applications, a weak plasmon damping reflected in the long relaxation time ($\tau$) is favorable, such that the dephasing of the optical polarization occurs slowly [42]. The complex dielectric function of SMO thin films demonstrates tunable plasmonic properties, as shown in Fig. 5. ENZ energies of 1.88 – 1.98 eV were obtained from $\varepsilon_1(\omega) = 0$ (Fig. 5(a) and its inset). The ENZ energies of the SMO epitaxial thin films are larger than those of $SrVO_3$ (1.30 eV) and $SrRuO_3$ (1.29 eV) and comparable to that of $SrNbO_3$ (1.90 eV) [43, 44]. The imaginary part of the dielectric function ($\varepsilon_2(\omega)$) (Fig. 5(b)) exhibits a strong optical transition at ~4.7 eV in addition to the prominent Drude absorption, originating from the interband transition between the O $2p$ and Mo $4d$ $t_{2g}$ states, as schematically shown in the inset of Fig. 5(b). $\omega_p$ calculated using Eq. 2 were 4.05 – 4.76 eV as shown in Table 1. The loss function (*LF*) was calculated using $LF = -\mathrm{Im}\,[\varepsilon^{-1}(\omega)] = \varepsilon_2(\omega)/(\varepsilon_1^2(\omega)+\varepsilon_2^2(\omega))$. The peak energy in the *LF* ($E(LF_{max})$) is attributed to plasmon excitation and is positioned at slightly higher energies (1.97 – 2.02 eV) than the ENZ energies for all the films (Fig. 5(c)). This is a signature of the damping of free carriers in the metal [40]. From those energy shifts, we can obtain $\tau$ as

$$\hbar/\tau = \sqrt{E(LF_{max})^2 - \left(\hbar\omega_{\varepsilon_1(\omega)=0}\right)^2}, \qquad (4)$$

resulting in $\tau = 1.12 - 1.63 \times 10^{-15}$ s, as summarized in Table 1. The $\tau$ values are observed to be in the same order but slightly larger than that of the SMO thin film on the silica glass substrate ($0.77 \times 10^{-15}$ s) [45].

Another characteristic time scale can be obtained by introducing the total plasmon dephasing time $T_2$ with the relationship $T_2 = 2\hbar/\Gamma$ [46]. Here, $\Gamma$ is the FWHM of the resonant *LF* peak. $T_2$ accounts for the overall plasmon damping and is given by the following equation,

$$\frac{1}{T_2} = \frac{1}{2T_1} + \frac{1}{T_2^*}, \qquad (5)$$

where $T_1$ is the relaxation time of the plasmon involving both radiative and nonradiative processes. $T_2^*$ is the pure dephasing time originating from other collisional mechanisms that alter the plasmon wave vector without varying its total energy. Because $T_2^*$ is negligible, the relaxation time originates mostly from $T_1$ defined as $T_2 \approx 2T_1$ [46, 47]. The $T_1$ values are found to be $5.48 – 6.68 \times 10^{-16}$ s for all the films, shorter than the $\tau$ values calculated using Eq. 4. This implies that there is another scattering mechanism that contributes to the dephasing of the plasmon in addition to the simple free electron scattering in this correlated metal.

**Table 1.** Comparison of the epsilon-near-zero (ENZ), plasma frequency ($\omega_p$), scattering time ($\tau$), the peak position in the loss function ($E(LF_{max})$), Lorentz broadening ($\Gamma$), and plasmon relaxation time ($T_1$) for all the SMO thin films.

| Fluence [J·cm$^{-2}$] | ENZ [eV] | $\hbar\omega_p$ [eV] | $\tau$ [fs] | $E(LF_{max})$ [eV] | $\Gamma$ [eV] | $T_1$ [fs] |
|---|---|---|---|---|---|---|
| 0.48 | 1.88 | 4.07 | 1.12 | 1.97 | 0.96 | 0.68 |
| 0.61 | 1.98 | 4.76 | 1.63 | 2.02 | 1.13 | 0.58 |
| 0.74 | 1.92 | 4.21 | 1.23 | 2.00 | 0.99 | 0.67 |
| 0.86 | 1.89 | 4.05 | 1.17 | 1.98 | 1.20 | 0.55 |

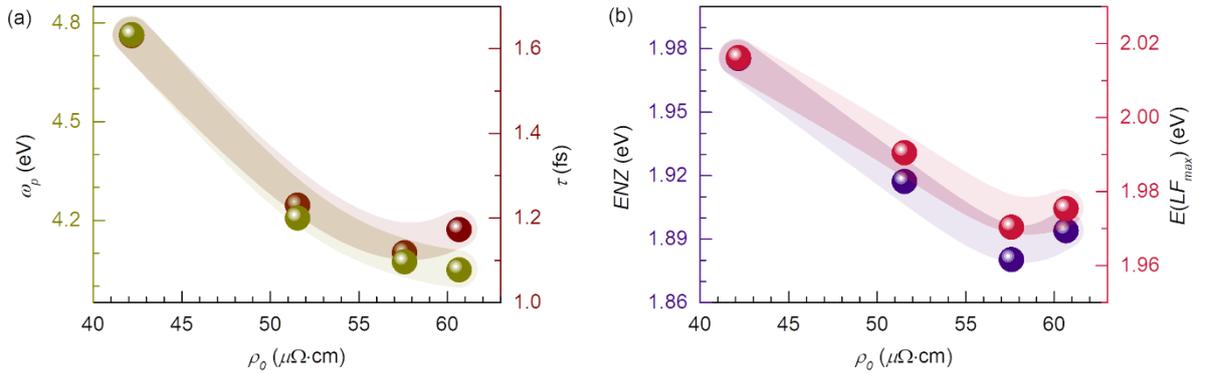

**Figure 6.** (a) The plasma frequency ($\omega_p$, left) and scattering time ($\tau$, right) as a function of $\rho_0$. (b) The epsilon-near-zero (ENZ, left) and energy of the peak occurred in the loss function ($E(LF_{max})$, right) as a function of $\rho_0$.

The structural quality modulation proves the effectiveness of modifying the plasmonic properties of the SMO thin films, as shown in Fig. 6. $\omega_p$, ENZ, and $E(LF_{max})$ shifted to higher energies for the film, whereas $\tau$ increased with a decrease in $\rho_0$. Here, $\rho_0$ is used to be the control parameter reflecting the structural quality as it increases with increasing FWHM as shown in Fig. 4(a). $\omega_p$ was observed to decrease from 4.76 to 4.05 eV (~15%), whereas the ENZ energies were tuned from 1.88 to 1.98 eV (~5%) with an increasing degree of electron-impurity scattering reflected in $\rho_0$. The observed variation in the ENZ energy was less than that in $\omega_p$. This was owing to the pronounced effect of electron-impurity scattering with an increase in $\rho_0$ followed by a decrease in mean free path and $\tau$. The $T$-independent electron-impurity scattering influences the degree of electron-electron interaction reflected in $T^*$ as shown in Fig. 4(b). The $T^*$ values were shown to be inversely proportional to the $\rho_0$. In SMO, there are two types of scattering mechanisms that govern the overall transport properties depending on the $T$ range, as discussed in Fig. 1, the electron-electron interaction below the $T^*$ exhibiting FL behavior and electron-phonon interaction above the $T^*$ displaying the non-FL behavior. Previously in the perovskite oxides, the tunability of the electron-electron correlation has been employed to shift $\omega_p$ to the infrared region by the enhancement of $m^*$ as shown in Fig. 1(a) [25, 48]. Meanwhile, the increase in the electron-impurity scattering leads to an increase of $\rho_0$ and $\tau$, shifting $\omega_p$ to a higher energy region. Here, the electron-impurity scattering was shown to significantly influence the overall transport and plasmonic behavior of SMO thin films in the whole $T$ range overcoming the contribution from both electron-electron interaction below $T^*$. Finally, our results corroborates the importance of tuning the electron scattering mechanism via the modulation of structural quality in obtaining the targeted plasmonic behavior in the visible region.

## 3. Conclusion

In summary, we have shown that the transport and plasmonic properties of SMO are highly dependent on the collective electron scattering mechanism governed by the structural quality of the film. This was demonstrated by the close correlation between $\rho_0$, $T^*$, and $\omega_p$. Higher-quality films result in a lower $\rho_0$ leading to an enhanced electron-electron interaction with higher $T^*$, and higher $\omega_p$, which is not expected by considering the electron-electron interaction alone. This proves the influence of $T$-independent electron-impurity scattering displayed by $\rho_0$ in defining the plasmonic properties represented in the ENZ energy and $\tau$. Finally, despite the consistent structural and transport properties, the delicate tunability of the electron scattering mechanism through structural quality control is highly important for realizing the plasmonic behavior of SMO in the visible region for viable plasmonic sensing applications.

**Acknowledgments**

This work was supported by the Basic Science Research Program through the National Research Foundation of Korea (NRF-2021R1A2C2011340, NRF-2022R1C1C2006723, and NRF-2022M3H4A1A04085306).

# Supporting Information

**Tunable electron scattering mechanism in plasmonic SrMoO3 thin films**

*Rahma Dhani Prasetiyawati, Seung Gyo Jeong, Chan Koo Park, Sehwan Song, Sungkyun Park, Tuson Park, and Woo Seok Choi*

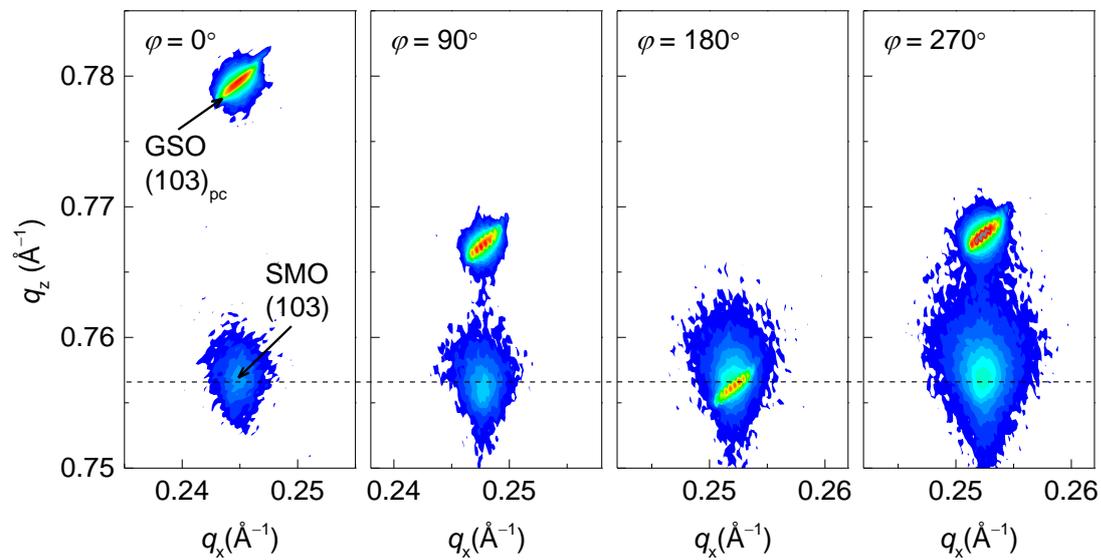

**Figure S1.** Reciprocal space maps of SMO thin film deposited on GSO substrate around (103)$_{pc}$ reflections at four different $\varphi$ angles of substrate orientation ($\varphi$ = 0°, 90°, 180°, and 270°) confirming the tetragonality of SMO film.

## S1. Atomic concentrations of the SMO thin films

The SMO epitaxial thin films revealed consistent $Mo^{4+}$ signals independent of the modification of the laser fluence, as depicted in Fig. S2. The oxidation state of Mo plays an essential role in determining the electronic structure and transport property of the SMO thin film by influencing the hybridization of Mo 4$d$ orbitals with O 2$p$ orbitals. Although the XPS for the Mo 3$d$ core level may indicate a multi-valent characteristic of Mo (Fig. S2(a)), most of the signals other than those originating from $Mo^{4+}$ are determined to be extrinsic. For example, the appearance of a metallic $Mo^0$ peak at the binding energies of 228.3 and 231.5 eV is generally caused by the over-reduction of Mo during the $Ar^+$-ion etching before the XPS measurement [1]. This etching process also oxidized $Mo^{4+}$ on the surface in $Mo^{6+}$ as $Mo^{6+}$ peaks appear at binding energies of 232.5 and 235.6 eV [2]. Nevertheless, the $Mo^{4+}$ state dominates the observed XPS spectra in the presence of screened $Mo^{4+}$ peaks ($Mo^{4+}_s$), at the lower energies (229.3 and 232.5 eV). The unscreened peaks ($Mo^{4+}_u$) are located at higher binding energies (231.0 and 234.2 eV) [3]. The absence of the $SrMoO_4$ phase in the XRD results further supports the consistent $Mo^{4+}$ valence state in the thin films. The Sr 3$d$ core level (Fig. S2(b)) shows the spin-orbit splitting of the Sr peaks into $3d_{5/2}$ and $3d_{3/2}$ peaks, as expected. Fig. S2(c) shows that the O 1$s$ XPS results are composed of three peaks, the two peaks around 532.0 eV correspond to the O 1$s$ contribution, and a peak centered at 530.4 eV originates from the Mo–O bond. The valence band spectra depicted in Fig. S2(d) confirm the Mo 4$d$ band crossing at the Fermi level, which indicates the metallic nature of the films. From the XPS analyses, we further assert that the atomic concentrations of all the films were rather similar as shown in Table S1, supporting the laser-fluence-independent lattice structures in the XRD results.

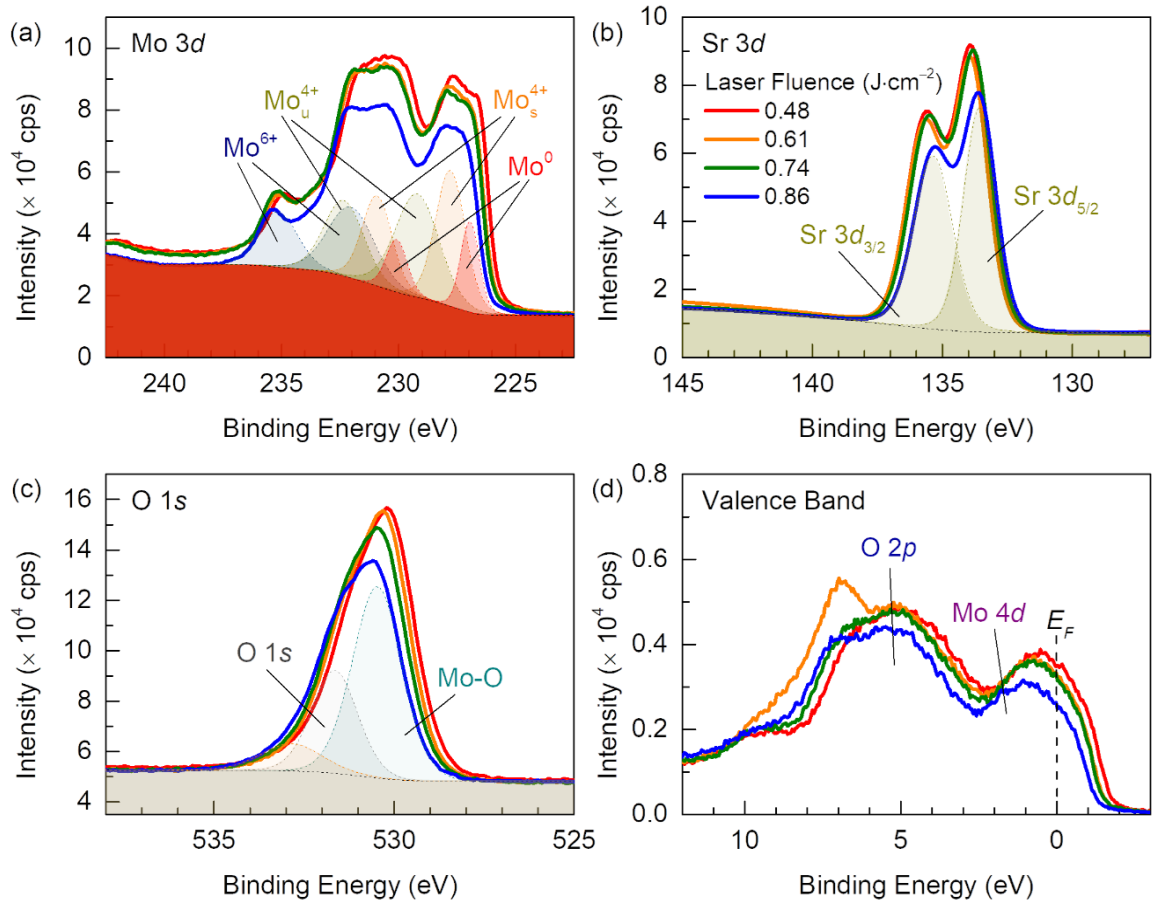

**Figure S2.** X-ray photoelectron spectra of SMO thin films in the (a) Mo 3*d*, (b) Sr 3*d*, (c) O 1*s*, and (d) valence band region. The oxidation state of Mo was confirmed to be 4+ with two components, screened peaks at low binding energies and unscreened peaks at higher binding energies.

**Table S1.** The cation concentration ratio (Sr/Mo) for all the SMO thin films.

| Fluence [J·cm$^{-2}$] | Sr/Mo Ratio |
|---|---|
| 0.48 | 0.91 ± 0.06 |
| 0.61 | 0.90 ± 0.05 |
| 0.74 | 0.93 ± 0.04 |
| 0.86 | 0.93 ± 0.02 |

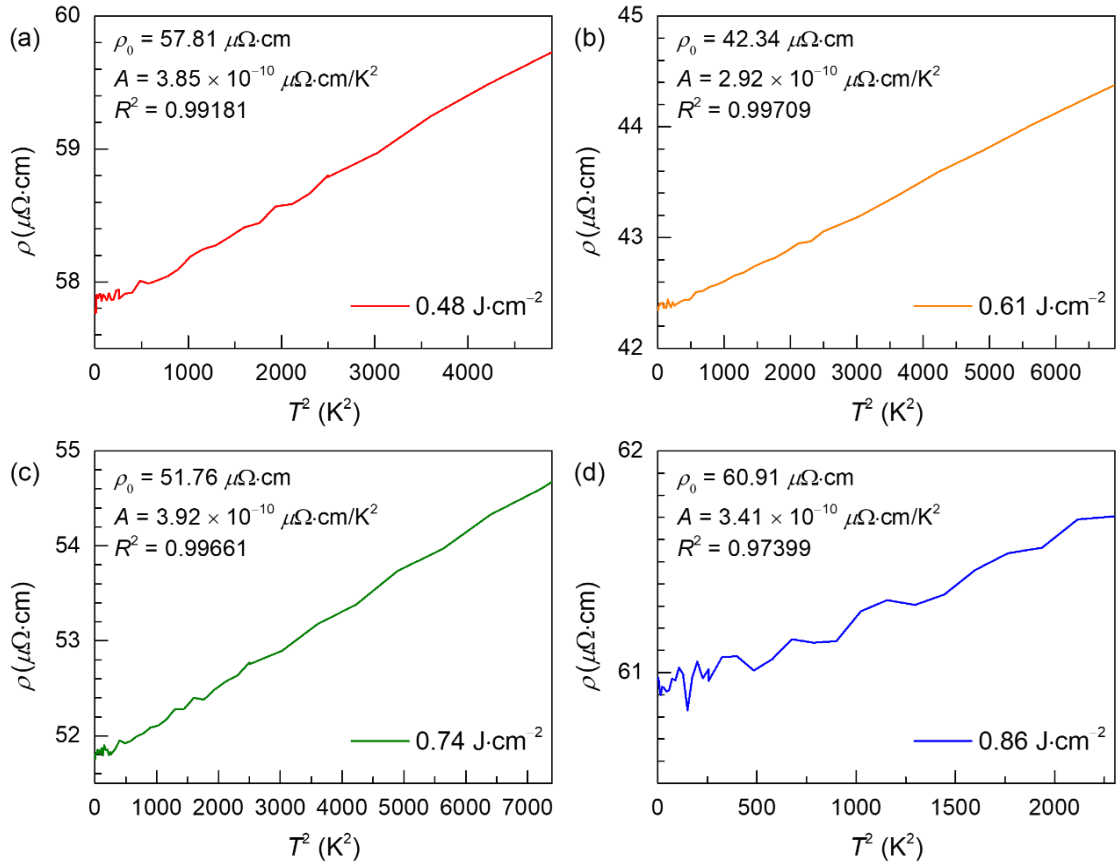

**Figure S3.** Fitting of $\rho$ to $T^2$-dependence below $T^*$ for SMO film grown using laser fluences of (a) 0.48, (b) 0.61, (c) 0.74, and (d) 0.86 J·cm$^{-2}$.

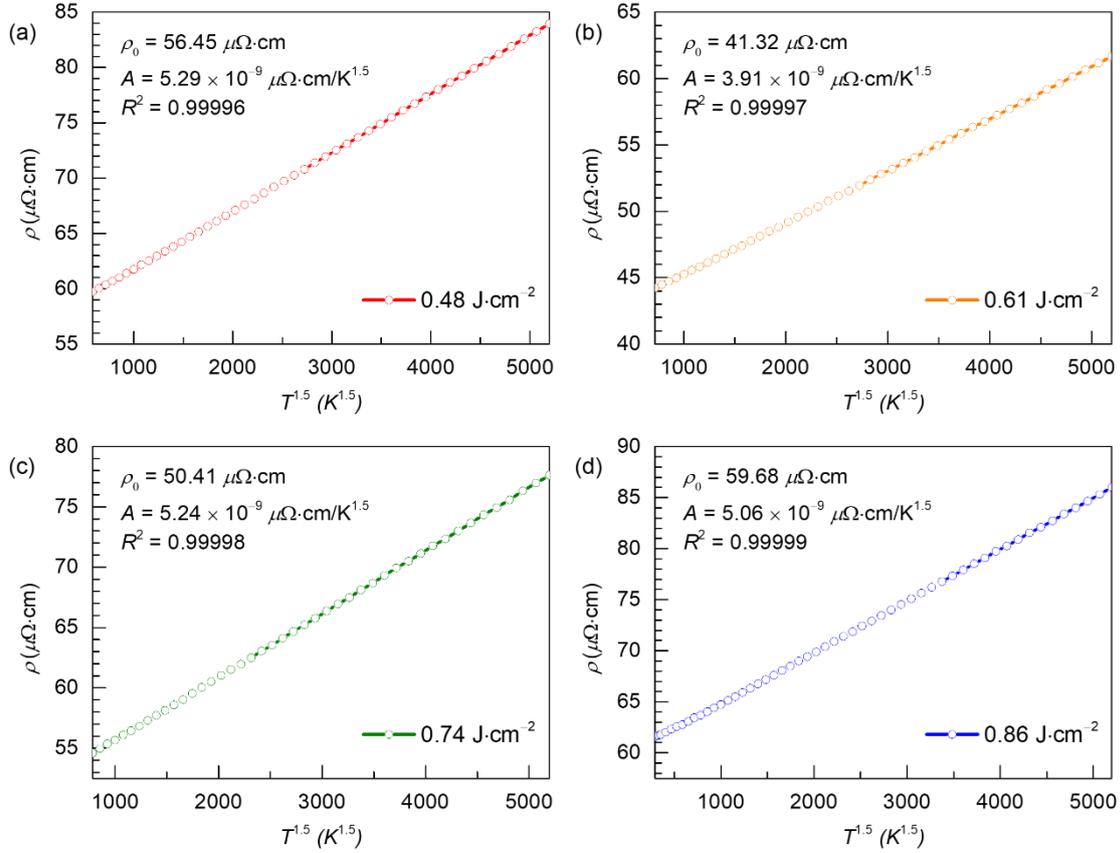

**Figure S4.** Fitting of $\rho$ to $T^{1.5}$-dependence above $T^*$ for SMO film grown using laser fluences of (a) 0.48, (b) 0.61, (c) 0.74, and (d) 0.86 J·cm$^{-2}$.

### S2. Hall measurement results of SMO thin films

From the Hall measurement, the magnetic field-dependent Hall resistances $R_H$ of SMO thin films measured at $T$ = 1.5, 5, 50, 100, 200, and 300 K exhibited linearity with the magnetic field as shown in Fig. S5(a). The $n$ and $\mu$ were also obtained for the SMO thin films (Fig. S5(b)). $n$ was found to be 2.41 – 2.99 × 10$^{22}$ cm$^{-3}$, close to the theoretical $n$ value whilst $\mu$ was found to be 3.09 – 3.38 cm$^2$ V$^{-1}$ s$^{-1}$ at room-$T$. $\mu$ increased with the decrease in $T$ up to 4.82 – 5.08 cm$^2$ V$^{-1}$ s$^{-1}$ at 5 K. The variation in the slope corresponding to an increase in $\mu$ with a decrease in $T$ approximately corresponds to $T^*$.

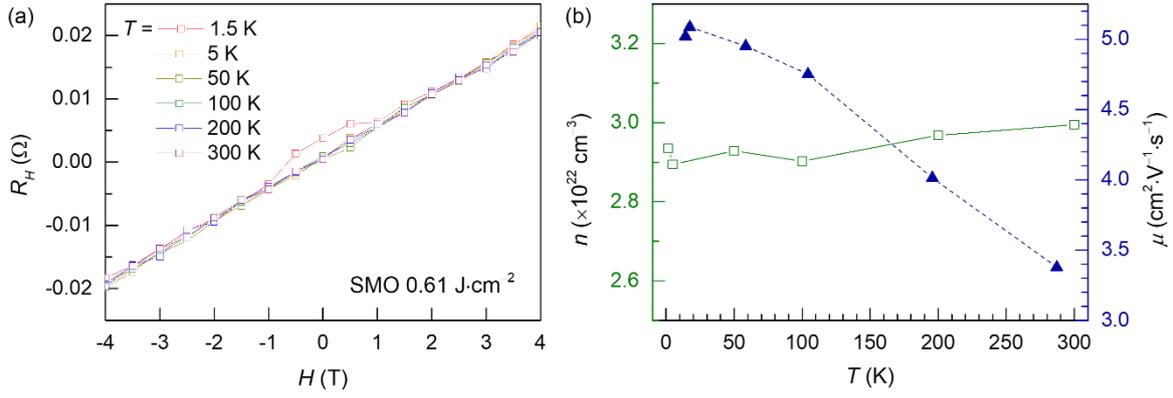

**Figure S5.** Hall measurement of SMO thin film grown using the laser fluence of 0.61 J·cm$^{-2}$. (a) The field-dependent Hall resistance $R_H$ of SMO thin film measured at temperatures of $T$ = 1.5, 5, 50, 100, 200, and 300 K. (b) $n$ (left) and $\mu$ (right) of SMO thin film as functions of $T$.

**S3. Thickness dependence of the structural and transport properties of SMO thin films**

The dependence of structural and transport properties of SMO thin films on the film thickness is shown in Fig. S6. Both the FWHM and $\rho_0$ increase with increasing film thickness (Fig. S6(a)). Yet, $\rho_0$ tends to be saturated when the film becomes even thicker. Meanwhile, *RRR* and $T^*$ values exhibit an opposite trend with an increase in film thickness (Fig. S6(b)). This can be attributed to the enhanced electron scattering by the impurities governed by the structural quality of the film. The increase in film thickness further suppresses the degree of electron-electron interaction resulting in a lower *RRR* and $T^*$ consistent with our findings.

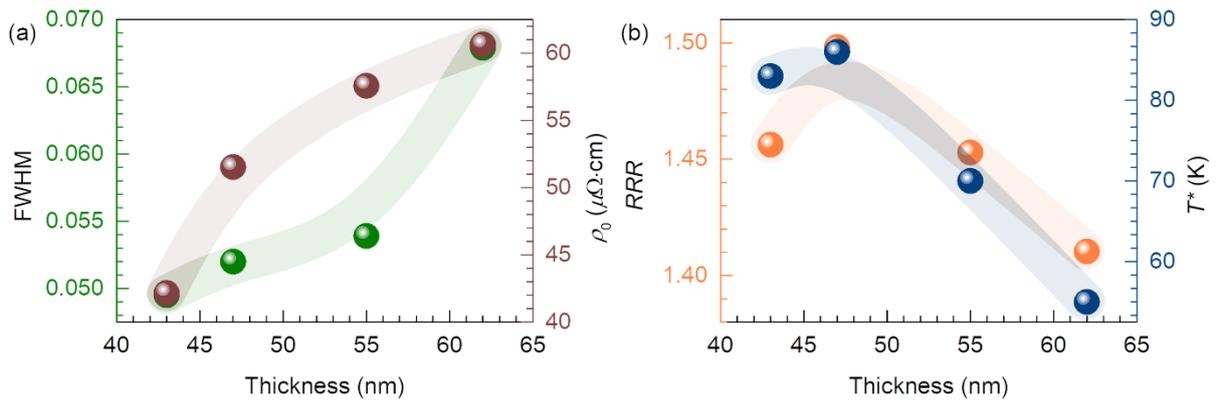

**Figure S6.** (a) The full-width-at-half-maximum (FWHM, left) and residual resistivity ($\rho_0$, right) of the SMO thin films as a function of the film thickness. (b) The residual resistivity ratio (*RRR*, left) and transition temperature ($T^*$, right) of SMO thin films as a function of the film thickness.

## S4. Power-law analysis results for the reported $T^*$ values of SMO

The recalculation of the reported $T^*$ values using the power-law analysis resulted in significant discrepancies between the calculated and reported $T^*$ values (Fig. S7). The recalculated values were 120, 80, and 65 K for single-crystal SMO, SMO on a SrTiO$_3$ (STO) thin film, and an SMO film on 10 unit-cell STO buffer layer/Ca$_2$Nb$_3$O$_{10}$ nanosheets/fused silica substrate, respectively [4–6]. These values differ significantly from the reported values. However, the trend of an increase in $T^*$ values with an increase in *RRR* is maintained. These results further corroborate the robustness of the power-scale analysis in accurately defining the $T^*$ value of the SMO system.

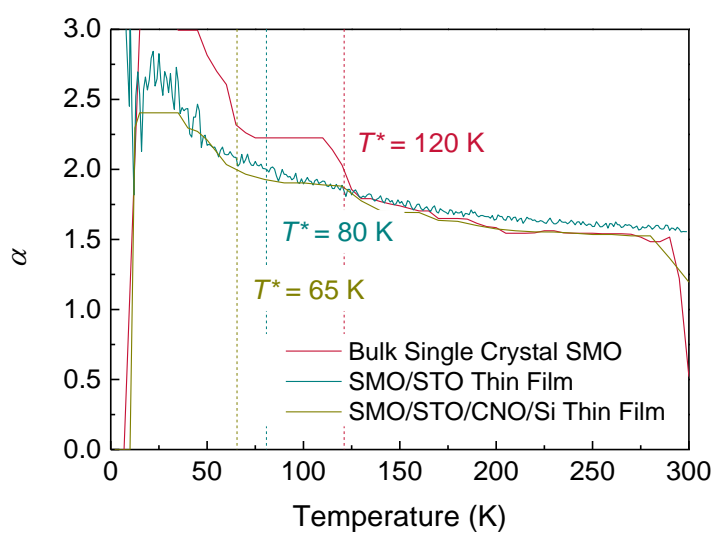

**Figure S7.** Calculated exponent $\alpha$ as a function of $T$ for single crystal SMO, SMO on SrTiO$_3$ thin film, and SMO/SrTiO$_3$ buffer/Ca$_2$Nb$_3$O$_{10}$ nanosheets/SiO$_2$ thin film.